7 files follow, each beginning with:

file.name-----------------------------------------------------------------------

the attached files are:

main.tex
aps.sty
aps10.sty
preprint.sty
revtex.sty
fig1.ps
fig2.ps

Separate the files, remove the file.name--- line from each, place
main.tex and the *.sty files in the same directory, and run Latex on
main.tex.  Print the PostScript files for the figures sepatately.

main.tex------------------------------------------------------------------------
\documentstyle[preprint,revtex]{aps}
\begin{document}
\draft
\begin{title}
Hadron Mass Predictions of the Valence Approximation\\ to Lattice QCD
\end{title}
\author{F. Butler, H. Chen, J. Sexton\cite{Trinity}, A. Vaccarino,\\
and D. Weingarten}
\begin{instit}
IBM Research \\
P.O. Box 218, Yorktown Heights, NY 10598
\end{instit}
\begin{abstract}
We evaluate the infinite volume, continuum limits of eight hadron mass
ratios predicted by lattice QCD with Wilson quarks in the valence
(quenched) approximation.  Each predicted ratio differs from the
corresponding observed value by less than 6\%.
\end{abstract}
\narrowtext

A key goal of the lattice formulation of QCD is to reproduce the masses
of the low-lying baryons and mesons. Lattice QCD mass predicitions for
the real world are supposed to be obtained from masses calculated with
finite lattice spacing and finite lattice volume by taking the limits of
zero spacing and infinite volume. In addition, since the algorithms used
for hadron mass calculations become progressively slower for small quark
masses, results are presently found with quark masses much larger than
the expected values of the up and down quark masses.  Predictions for
the masses of hadrons containing up and down quarks then require a
further extrapolation to small quark mass.  We report here mass
predictions combining all three extrapolations for Wilson quarks in the
valence (quenched) approximation. This approximation may be viewed as
replacing the momentum and frequency dependent color dielectric constant
arising from quark-antiquark vacuum polarization with it zero-momentum,
zero-frequency limit and might be expected to be fairly reliable for
low-lying baryon and meson masses~\cite{Weingar81}.

To our knowledge there have been no previous systematic attempts to
extrapolate hadron masses to physical quark mass, zero lattice spacing
and infinite volume. For a review of lattice QCD mass calculations see
Ref.~\cite{Toussaint}.

Our main result consists of a prediction of eight different hadron mass
ratios.  Each of the predicted ratios differs from experiment by less
than 6\%. In each case, the error is less than a factor of 1.6
multiplied by the corresponding statistical uncertainty.  We believe it
is reasonable to take these results as quantitative confirmation of the
mass predictions both of QCD and of the valence approximation. It seems
unlikely to us that the valence approximation would agree with
experiment for eight different mass ratios yet differ significantly from
QCD's predictions including the full effect of quark-antiquark vacuum
polarization.

Following Refs.~\cite{Fermilab,Lepage}, we also determine the continuum
coupling constant, $g^{(0)}_{\overline{ms}}$, from the lattice coupling
constant, $g_{lat}$, and from $g^{(0)}_{\overline{ms}}$ determine
$\Lambda^{(0)}_{\overline{ms}}$.  Two independent calculations of
$\Lambda^{(0)}_{\overline{ms}}$, done by rather different
methods~\cite{Fermilab,Wupper} lie within the 4\% statistical
uncertainty of our continuum, infinite-volume result for
$\Lambda^{(0)}_{\overline{ms}}$.  Values we obtain for the rho mass at
finite lattice spacing, measured in units of inverse lattice spacing,
depend on $g^{(0)}_{\overline{ms}}$ and $\Lambda^{(0)}_{\overline{ms}}$
as predicted by asymptotic scaling.  This result tends to support the
reliability of our extrapolation of masses to the continuum limit.

In addition to comparing the valence approximation to QCD with
experiment, a goal of the present work is to develop technology which
might be useful in extrapolating results of the full theory to physical
quark mass, infinite volume, and zero lattice spacing.

The calculations described here were done on the GF11 parallel computer
at IBM Research \cite{Weingar90} and took approximately one year to
complete.  GF11 was used in configurations ranging from 384 to 480
processors, with sustained speeds ranging from 5 Gflops to 7 Gflops.
With the present set of improved algorithms and 480 processors, these
calculations could be repeated in less than 4 months.

Table~\ref{tab:lattices} lists the lattice sizes and parameter values
for which hadron propagators were evaluated.  We chose periodic boundary
conditions in all directions for both gauge fields and quark fields.
Gauge configuations were generated using a version of the
Cabbibo-Marinari-Okawa algorithm, with the number of sweeps skipped
between configurations and total count of configurations as given in the
table.  A variety of correlation tests showed all of the configurations
on which propagators were evaluated were statistically independent.

For the $8^3 \times 32$ lattice at $\beta$ of 5.7 we used point sources
and sinks in the quark propagators. For all other lattices and $\beta$,
each gauge configuration was transformed to lattice Coulomb gauge and
quark propagators were then found for gaussian extended sources and for
point sinks and four different sizes of gaussian sinks\cite{Smear}.  The
mean squared radius of the gaussian source in all cases was 6, in
lattice units.  The mean squared radii of the gaussian sinks ranged from
1.5 to 24, in lattice units.  On the lattice $24^3 \times 32$ at $\beta$
of 5.7, eight independent gaussian sources were placed on the source
hyperplane, each multiplied by a random cube root of 1 to cancel cross
terms between the propagation of different sources for both baryon and
meson propagators.

Quark propagators were constructed using the conjugate gradient
algorithm for the $8^3 \times 32$ lattice at $\beta$ of 5.7, using
red-black preconditioned conjugate gradient for the other lattices at
$\beta$ of 5.7 and 5.93, and using a red-black preconditioned minimum
residual algorithm at $\beta$ of 6.17.  At the largest hopping constant
values at $\beta$ of 5.7 and 5.93, preconditioning the conjugate
gradient algorithm increased its speed by a factor of 3, and at the
largest hopping constant values at $\beta$ of 6.17, the change from
conjugate gradient to the minimum residual algorithm yielded an
additional factor of 2 in speed.  The convergence criterion used in all
cases was equivalent to the requirement that effective pion, rho,
nucleon and delta masses evaluated between successive pairs of time
slices must be within 0.2\% of their values obtained on propatagors run
to machine precision.

Hadron masses were determined by fits to hadron propagators constructed
from the quark propagators. The pion mass for all values of the hopping
constant $k$, and the rho, nucleon and delta masses for all but the
three largest values of $k$ at each $\beta$, were obtained from the
propagators for a point sink.  At the largest three $k$ values, the rho,
nucleon and delta baryon masses were found by simultaneously fitting a
single mass value to the propagators for a point sink and for gaussian
sinks with mean squared radius of 1.5 and 6.  The statistical errors for
all fits were determined by the bootstrap method. A more detailed
discussion of our fits and error analysis will be given
elsewhere~\cite{Toappear}.

Comparing hadron masses in lattice units between the $8^3 \times 32$ and
$16^3 \times 32$ lattices at $\beta$ of 5.7 for $k$ up to 0.1650 showed
no statistically significant differences. Comparing $16^3 \times 32$ and
$24^3 \times 32$ for $k$ up to 0.1675 showed no statistically
significant differences in the pion mass.  For the rho, nucleon and
delta, marginally significant differences were found at the largest $k$.
Percentage changes in mass going from $16^3 \times 32$ to $24^3 \times
32$ are given in Table~\ref{tab:voldep}.  Although some of the changes
shown in Table~\ref{tab:voldep} take smaller values if we use different
procedures to determine hadron masses from hadron propagators, none
become larger \cite{Toappear}.  Thus a conservative interpretation of
the changes in Table~\ref{tab:voldep} is to view them primarily as upper
bounds on volume dependence. It appears quite likely that for the range
of $k$, $\beta$ and lattice volume we have examined, the errors in
valence approximation hadron masses due to calculation in a finite
volume $L^3$ are bounded by an expression of the form $C e^{- L/R}$,
with a coefficient $R$ of the order of the radius of a the hadron's wave
function. At $\beta$ of 5.7 for the $k$ we considered, $R$ is thus
typically 3 lattice units.  We therefore expect that the differences
between masses on a $16^3$ volume and those on a $24^3$ volume are
nearly equal to the differences between $16^3$ and true infinite volume
limiting values.

At the largest $k$ on each lattice, except $8^3 \times 32$, the ratio
$m_{\pi} / m_{\rho}$ is close to 0.5. For the lattice $8^3 \times 32$,
this ratio is 0.691. These values of $m_{\pi} / m_{\rho}$ are
significantly above the experimentally observed value of 0.179 for
charge averaged $m_{\pi}$ and $m_{\rho}$. To produce mass predictions
for hadrons containing only light quarks our data has to be extrapolated
to larger $k$ or, equivalently, to smaller quark mass.  We did not
calculate hadron masses directly at larger $k$ both because the
algorithms we used to find quark propagators became too slow and because
the statistical errors we found in trial calculations became too large.

For each lattice except $8^3 \times 32$, we extrapolated hadron mass
values down to small quark mass.  To do this we first determined the
$k_{crit}$ at which $m_{\pi}$ becomes 0.  As expected from a naive
application of PCAC, $(m_{\pi} a)^2$ turned out to be roughly a linear
function of $1/k$ over the entire range of $k$ considered on each
lattice, and for the three largest $k$ on each lattice, $(m_{\pi} a)^2$
fit a linear function of $1/k$ quite well.  From these fits we then
found $k_{crit}$ for each lattice and $\beta$.  Defining the quark mass
in lattice units, $m_q a$, to be $1/(2 k) - 1/(2 k_{crit})$, we found
$m_{\rho} a$, $m_N a$ and $m_{\Delta} a$ to be roughly linear functions
of $m_q a$ over the entire range of $k$ considered on each lattice and
quite close to linear functions at the three smallest $m_q$
(corresponding to the three largest $k$).  Figure~\ref{fig:massfit}
shows $m_{\pi}^2$, $m_{\rho}$, $m_N$, and $m_{\Delta}$, for the lattice
$30 \times 32^2 \times 40$ at $\beta$ of 6.17, as functions of $m_q$.
The lines in Figure~\ref{fig:massfit} are fits to each data set at the
three smallest values of $m_q$. For convenience, we show all hadron
masses in units of the physical rho mass, $m_{\rho}(m_n)$, given by
$m_{\rho}$ evaluated at the ``normal'' quark mass $m_n$ which produces
the physical value of $m_{\pi}/m_{\rho}$. The quark mass $m_q$ in
Figure~\ref{fig:massfit} is shown in units of the strange quark mass
$m_s$, the determination of which will be discussed below.  The fits
shown in Figure~\ref{fig:massfit} appear to be reasonably good and
provide, we believe, a reliable method for extrapolating hadron masses
down to light quark masses.  Fits comparable to those shown were
obtained for the nucleon, rho and delta baryon on all the lattices we
considered except $8^3 \times 32$.

Using a version of the Gell-Mann-Okubo mass formula, the accuracy of
linear extrapolation in quark mass can be checked with observed hadron
masses.  For a rho composed of a quark with mass $m_1$ and an antiquark
with mass $m_2$, Figure~\ref{fig:massfit} suggests $m_{\rho} = \alpha_1
m_1 + \alpha_2 m_2 + \beta$.  Charge conjugation invariance then gives
$\alpha_1 = \alpha_2$.  It follows that the k-star, which is a rho with
$m_1 = m_s$ and $m_2 = m_n$, will have the same mass as a rho composed
of a single type of quark with $m_1 = m_2 = (m_s + m_n)/2$.  In the
valence approximation the phi is a rho with $m_1 = m_2 = m_s$. The
linear relation $m_{\rho} = \alpha (m_1 + m_2) + \beta$ then permits the
rho mass to be extrapolated from the masses of k-star and phi.  The
extrapolated rho mass obtained from observed k-star and phi masses lies
below the observed rho mass by 0.53 \%.  Similar extrapolations can be
made to determine the nucleon mass from the observed masses of its
strange partners and to determine the delta baryon mass from its strange
partners.  The extrapolated nucleon mass is larger than experiment by
1.38 \%, and the extrapolated delta baryon mass is large by 0.81 \%.

The relations discussed in the preceding paragraph can also be used to
determine strange hadron masses from the masses we have calculated for
hadrons composed of a single species of heavy quark.  Fitting the kaon
to the pion mass at a quark mass of $(m_s + m_n)/2$ gives the value for
$m_s$ mentioned above. With $m_s$ and $m_n$ thus determined, the ratios
of eight different hadron mass combinations to the physical rho mass
follow from our data with no additional free parameters.

These ratios we extrapolated to zero lattice spacing with the physical
lattice volume nearly held fixed.  For Wilson fermions the leading
lattice spacing dependence in mass ratios is expected to be linear in a.
Figure~\ref{fig:nuc_extrap} shows a linear fit of our data for $m_N /
m_{\rho}$ to the rho mass, at physical quark mass, measured in lattice
units, $m_{\rho} a$.  The horozontal axis may also be interpreted as the
lattice spacing $a$ measured in units of $1/m_{\rho}$.  The vertical bar
at $m_{\rho} a$ of 0 is the extrapolated prediction's uncertainty,
determined by the bootstrap method. The dot at $m_{\rho} a$ of 0 is the
experimental value of $m_N / m_{\rho}$.  The three data points in
Figure~\ref{fig:nuc_extrap} are for the lattices $16^3 \times 32$, $24^3
\times 36$ and $30 \times 32^2 \times 40$.  The values of $\beta$ for
these lattices were chosen so that the physical volume in each case is
nearly the same.  For lattice period L, the quantity $m_{\rho} L$ is
respectively, 9.08 $\pm$ 0.13, 9.24 $\pm$ 0.19 and, averaged over three
directions, 8.67 $\pm$ 0.12.

The continuum ratios we found in finite volume were then extrapolated to
infinite volume. This was done by using the differences between mass
ratios found on the lattice $16^3 \times 32$, at $\beta$ of 5.7, and
mass ratios found on the lattice $24^3 \times 32$, at $\beta$ of 5.7, as
finite lattice spacing approximations to the differences between
continuum mass ratios in a box with period having $m_{\rho} L$ of 9 and
continuum mass ratios in infinite volume.  The error in this procedure
can be estimated to be about 1 \% as follows.  All of the finite volume
extrapolated zero lattice spacing mass ratios which we obtained were
within 20 \% of their values on the lattice $16^3 \times 32$ at $\beta$
of 5.7. Moreover, as we argued earlier, the changes in masses we found,
at $\beta$ of 5.7, between $16^3 \times 32$ and $24^3 \times 32$ should
be nearly the same as corresponding changes between $16^3 \times 32$ and
inifinite volume.  Combining these two pieces of information, we expect
that with a relative error of 20 \% or less, the changes we found in
mass ratios between $16^3 \times 32$ and $24^3 \times 32$ at $\beta$ of
5.7 should be the same as the changes between continuum mass ratios in a
box with period having $m_{\rho} L$ of 9 and corresponding continuum
ratios in infinite volume.  Since the changes we found in mass ratios,
extrapolated to physical quark mass, between $16^3 \times 32$ and $24^3
\times 32$ are all less than 5 \%, the overall error in using these
differences as estimates of corresponding continuum differences between
$m_{\rho} L$ of 9 and infinite volume should of the order of 20 \% of
5\%, which is 1 \%.

Eight different hadron mass ratios, extrapolated to zero lattice spacing
with $m_{\rho} L$ fixed at 9, and then extraplated to infinite volume
are shown in Table~\ref{tab:results}.  All eight infinite volume
continuum predictions differ from experiment by less than 6\% and less
than 1.6 standard deviations.  The central values of the infinite volume
ratios shown in Table~\ref{tab:results} are marginally closer to
experiment than the finite volume ratios.  We believe the main
significance of the infinite volume numbers shown in
Table~\ref{tab:results}, however, is that their error bars include the
uncertainty in estimating infinite volume ratios from finite volume.
Variations of our mass fitting procedure which decrease some of the
volume dependence shown in Table~\ref{tab:voldep} do not produce
statistically significant changes in the numbers shown in
Table~\ref{tab:results}.  The errors on all quantities in this table
were found by the bootstrap method.

Values of $\Lambda^{(0)}_{\overline{ms}}$, determined following
Ref.~\cite{Fermilab}, give $\Lambda^{(0)}_{\overline{ms}} / m_{\rho}$
which vary by only one standard deviation over the three lattices used
for extrapolation. Thus the rho mass follows asymptotic scaling in
$g^{(0)}_{\overline{ms}}$.  The continuum and infinite volume continuum
limits of $\Lambda^{(0)}_{\overline{ms}} / m_{\rho}$ are shown in
Table~\ref{tab:results}.  For the observed value of
$\Lambda^{(0)}_{\overline{ms}} / m_{\rho}$ we have inserted the
calculated results of Refs.~\cite{Fermilab,Wupper}.

The authors are grateful to Mike Cassera and Dave George for their work
in putting GF11 into operation, to Chi Chai Huang, of Compunetix Inc.,
for his contributions to bringing GF11 up to full power and to its
continued maintenance, and to Ed Nowicki and Molly Elliot for their work
on GF11's disk software.

\figure{ For a $30 \times 32^2 \times 40$ lattice at $\beta$ of 6.17,
$m_{\pi}^2$, $m_{\rho}$, $m_N$ and $m_{\Delta}$, in units of the
physical rho mass $m_{\rho}(m_n)$, as functions of the quark mass $m_q$,
in units of the strange quark mass $m_s$. The symbol at each point
is larger than the error bars.\label{fig:massfit}}

\figure{$m_N/m_{\rho}$ as a function of the lattice spacing $a$, in units of
$1/m_{\rho}$.  The straight line is an extraploation to zero lattice
spacing, the error bar at zero lattice spacing is the uncertainty in the
extrapolated ratio, and the point at zero lattice spacing is the
observed value. \label{fig:nuc_extrap}}

\begin{table}
\begin{tabular}{r@{}llllr}     \hline
 \multicolumn{2}{c}{lattice}          & $\beta$ & k     & skip & count\\
 \hline
 $8^3$ & $\:\times \: 32$ & 5.7 & 0.1400 - 0.1650 & 1000 & 2349\\
 $16^3$ & $\:\times \: 32$ & 5.7 & 0.1400 - 0.1550 & 2000 & 47\\
       &          &     & 0.1600 - 0.1675 & 2000 & 219\\
 $24^3$ & $\:\times \: 32$ & 5.7 & 0.1600 - 0.1675 & 4000 & 92\\
 $24^3$ & $\:\times \: 36$ & 5.93 & 0.1543 - 0.1581 & 4000 & 217\\
 $30 \times 32^2$ & $\:\times \: 40$ & 6.17 & 0.1500 - 0.1532 & 6000 & 219\\
 \hline
\end{tabular}
\caption{Configurations analyzed.}
\label{tab:lattices}
\end{table}

\begin{table}
\begin{tabular}{llr@{$\pm$}l}     \hline
 particle          & k     & \multicolumn{2}{c}{change}\\ \hline
 pion & all    & $ <1.2$ & 1.6 \%\\
 rho  & 0.1675 & $-3.4$ & 1.4 \%\\
 nucleon & 0.16625 & $-4.4$ & 1.7 \%\\
         & 0.1675 & $-4.6$ & 2.2 \%\\
 delta   & 0.1675 & $-4.7$ & 2.6 \%\\ \hline
\end{tabular}
\caption{Changes in mass from a lattice $16^3 \times 32$ to
a lattice $24^3 \times 32$ at $\beta$ of 5.7.}
\label{tab:voldep}
\end{table}

\begin{table}
\caption{
Calculated values of hadron mass ratios at physical quark masses,
extrapolated to zero lattice spacing in finite volume, then corrected to
infinite volume, compared with observed values, and with calculations of
Refs.~\cite{Fermilab,Wupper} for $\Lambda^{(0)}_{\overline{ms}} /
m_{\rho}$. The mass difference $\Delta m$ is $m_{\Xi} + m_{\Sigma} -
m_N$.}
\begin{tabular}{cr@{}l@{$\pm$}lr@{}l@{$\pm$}lr@{}l}
 \hline
 ratio & \multicolumn{3}{c}{finite volume} &
 \multicolumn{3}{c}{infinite volume}
  & \multicolumn{2}{c}{observed}\\ \hline

 $m_{K^*} / m_{\rho}$ & 1 & .149 & 0.010 &  1 & .167 & 0.016 & 1 & .164\\
 $m_{\Phi} / m_{\rho}$ & 1 & .297 & 0.019 &  1 & .333 & 0.032 & 1 & .327\\
 $m_{N} / m_{\rho}$ & 1 & .285 & 0.070 & 1 & .219 & 0.105 & 1 & .222\\
 $\Delta m / m_{\rho}$ & 1 & .867 & 0.046 & 1 & .930 & 0.073 & 2 & .047\\
 $m_{\Delta} / m_{\rho}$ & 1 & .628 & 0.075  &  1 & .595 & 0.111 & 1 & .604\\
 $m_{\Sigma^*} / m_{\rho}$ & 1 & .813 & 0.051 & 1 & .821 & 0.075 & 1 & .803\\
 $m_{\Xi^*} / m_{\rho}$ & 2 & .013 & 0.052  &  2 & .063 & 0.067 & 1 & .996\\
 $m_{\Omega} / m_{\rho}$ & 2 & .206 & 0.058 & 2 & .298 & 0.098 & 2 & .177\\
 $\Lambda^{(0)}_{\overline{ms}} / m_{\rho}$ & 0 & .305 & 0.008 &
0 & .319 & 0.012 & 0 & .305 $\pm$ 0.018\\
\multicolumn{7}{c}{} & 0 & .320 $\pm$ 0.007\\ \hline
\end{tabular}
\label{tab:results}
\end{table}

\end{document}